\documentclass[11pt]{article}
\usepackage[brazil]{babel}
\usepackage{url}
\usepackage{amssymb,amsxtra}
\usepackage{amsmath,color,graphicx}
\usepackage{lineno}
\usepackage{graphicx}
\RequirePackage{geometry}
\usepackage{indentfirst}

\begin{document}

\def\chaptername{}
\def\contentsname{Sum\'{a}rio}
\def\listfigurename{Figuras}
\def\listtablename{Tabelas}
\def\abstractname{Resumo}
\def\appendixname{Ap\^{e}ndice}
\def\refname{\large Refer\^{e}ncias bibliogr\'{a}ficas}
\def\bibname{Bibliografia}
\def\indexname{\'{I}ndice remissivo}
\def\figurename{\small Figura~}
\def\tablename{\small Tabela~}
\def\pagename{\small Pag.}
\def\seename{veja}
\def\alsoname{veja tamb\'em}
\def\na{-\kern-.4em\raise.8ex\hbox{{\tt \scriptsize a}}\ }
\def\pa{\slash \kern-.5em\raise.1ex\hbox{p}\ }
\def\ro{-\kern-.4em\raise.8ex\hbox{{\tt \scriptsize o}}\ }
\def\no{n$^{\underline{\rm o}}$}
\setcounter{tocdepth}{3}

\clearpage
\pagenumbering{arabic}

\thispagestyle{empty}
\parskip 8pt

\vspace*{0.2cm}
\begin{center}

\ \\

{\huge \bf \`{A}s vezes, solucionar uma crise depende de ampli\'{a}-la: a contribui\c{c}\~{a}o de Louis de Broglie \`{a} F\'{\i}sica Qu\^{a}ntica}\\
\vspace*{1.0cm}


{\Large \bf \it Francisco Caruso;$^{\,1}$ Vitor Oguri$^{\,2}$}\\[2.em]

{{$^{1}$ Centro Brasileiro de Pesquisas F\'{\i}sicas, Coordena\c{c}\~{a}o de F\'{\i}sica de Altas Energias, 22290-180, Rio de Janeiro, RJ, Brazil.}}

{{$^{2}$ Universidade do Estado do Rio de Janeiro, Instituto de F\'{\i}sica Armando Dias Tavares, 20550-900, Rio de Janeiro, RJ, Brasil.}}

\vfill
\end{center}

\noindent \textbf{Resumo}

Faz-se uma breve reconstru\c{c}\~{a}o hist\'{o}rica de alguns pontos que, de algum modo, contribu\'{\i}ram para o trabalho seminal de Louis de Broglie. Em particular, enfatiza-se a relev\^{a}ncia de sua tese de doutorado, principalmente por seu valor epistemol\'{o}gico, ao ampliar a crise que havia sido introduzida na descri\c{c}\~{a}o da radia\c{c}\~{a}o pelo \textit{quantum} de Planck, abrindo caminho para sua solu\c{c}\~{a}o.

\noindent \textbf{Palavras-chave:} Louis de Broglie; F\'{\i}sica Qu\^{a}ntica; Hist\'{o}ria da F\'{\i}sica.

\vspace*{0.7cm}





\newpage


\begin{flushright}
\begin{minipage}{8.5cm}
\baselineskip=10pt {\small
\textit{A hist\'{o}ria da ci\^{e}ncia mostra que o progresso da ci\^{e}ncia tem sido constantemente dificultado pela influ\^{e}ncia ti\-r\^{a}nica de certas concep\c{c}\~{o}es que finalmente passaram a ser consideradas dogmas. Por essa raz\~{a}o, \'{e} apropriado submeter periodicamente a um exame muito minucioso princ\'{\i}pios que acabamos por admitir sem mais nenhuma discuss\~{a}o. }}
\smallskip

\hfill Louis de Broglie
\end{minipage}
\end{flushright}


\section{Uma curiosidade}
\paragraph*{}

Poucas teses de doutorado tiveram um impacto t\~{a}o marcante -- e at\'{e} mesmo revolucion\'{a}rio -- no desenvolvimento da F\'{\i}sica Moderna quanto a do franc\^{e}s Louis de Broglie.~Sua ori\-gi\-nalidade, como sugere o t\'{\i}tulo do artigo, est\'{a}, n\~{a}o propriamente na solu\c{c}\~{a}o da crise introduzida pela quan\-tiza\c{c}\~{a}o da radia\c{c}\~{a}o de corpo negro de Max Planck~\cite{Bassalo-Planck}, mas na amplia\c{c}\~{a}o do dom\'{\i}nio da crise, que se convencionou chamar de \textit{dualidade onda-part\'{\i}cula}, tamb\'{e}m para a mat\'{e}\-ria~\cite{Caruso-Oguri}.

Sua tese foi apresentada \`{a} Faculdade de Ci\^{e}ncias de Paris, em 1924, dois anos antes da formula\c{c}\~{a}o da Mec\^{a}nica Ondulat\'{o}ria de Erwin Schr\"{o}dinger, ocasi\~{a}o em que houve certo embara\c{c}o por parte dos professores que iriam julg\'{a}-la, uma vez que a mesma fugia aos c\^{a}\-nones tradicionais da F\'{\i}sica.
\begin{figure}[htbp]
\centerline{\includegraphics[width=7.5cm]{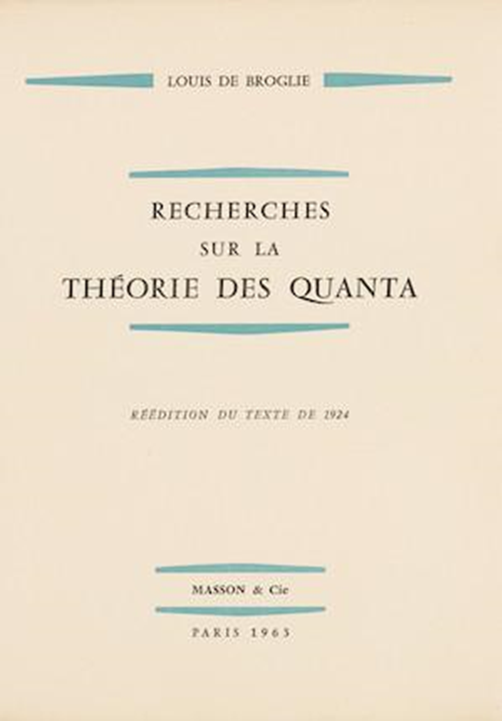}}
\caption{\baselineskip=8pt{\small Capa da reedi\c{c}\~{a}o, como livro, da tese de Louis de Broglie, \textit{Pesquisas sobre a Teoria dos Quanta}, 1963.}}
\label{capa_tese}
\end{figure}

Assim, ela foi encaminhada ao f\'{\i}sico franc\^{e}s Paul Langevin, para que ele desse seu parecer. De imediato, ele enviou uma c\'{o}pia ao seu amigo Albert Einstein que, por sua vez, pediu ao tamb\'{e}m f\'{\i}sico alem\~{a}o Max Born uma opini\~{a}o s\'{e}ria sobre a mesma, escrevendo-lhe: -- \textit{Leia isso! Embora pare\c{c}a ter sido escrita por um louco, est\'{a} escrito corretamente}~\cite{Bassalo-de-Broglie}.

Ao devolver a referida tese a Langevin, Einstein lhe disse que podia aprov\'{a}-la, j\'{a} que a mes\-ma continha muitas descobertas importantes. Dessa forma, Langevin foi o orientador formal da tese, defendida na Sorbonne, com o t\'{\i}tulo \textit{Recherches sur la th\'{e}orie des quanta}~\cite{These}, a qual foi apresentada e aprovada em 25 de novembro de 1924, quando de Broglie tinha 32 anos. Integravam a banca: Jean Baptiste Perrin, Charles-Victor Mau\-guin, mineralogista, e \'{E}lie Joseph Cartan, matem\'{a}tico, al\'{e}m do pr\'{o}prio Langevin.

\section{Os primeiros interesses}
\paragraph*{}

Em 1920, conforme declarou em sua \textit{Nobel Lecture}~\cite{Nobel-Lecture}, Louis de Broglie ficou fascinado pelos mist\'{e}rios escondidos na estrutura qu\^{a}ntica da mat\'{e}ria e da radia\c{c}\~{a}o, decorrentes do ``estranho conceito'' de \textit{quantum}. Entre essa data e o ano que precedeu a defesa de sua tese, seus interesses cient\'{\i}ficos orbitaram entre o papel do \textit{quantum} de Planck, os raios X, os espectros e o modelo de Bohr, podendo ser aferido pelos 12 trabalhos por ele publicados (todos no \textit{Comptes Rendus de l'Acad\'{e}mie des Sciences}) envolvendo temas sobre a radia\c{c}\~{a}o e sobre a mat\'{e}ria.~Foram eles: Sobre o c\'{a}lculo das frequ\^{e}ncias limites de absor\c{c}\~{a}o $K$ e $L$ dos elementos pesados~\cite{Paper-de-Broglie-1}, Sobre a absor\c{c}\~{a}o dos raios de R\"{o}ntgen pela ma\-t\'{e}\-ria~\cite{Paper-de-Broglie-2}, Sobre o modelo do \'{a}tomo de Bohr e os espectros corpusculares (com o irm\~{a}o Mau\-rice)~\cite{Paper-de-Broglie-3}, Sobre a estrutura eletr\^{o}nica dos \'{a}tomos pesados (com o qu\'{\i}mico Dauvillier)~\cite{Paper-de-Broglie-4}, Sobre a distribui\c{c}\~{a}o dos el\'{e}trons dentro dos \'{a}tomos pesados (com Dauvillier)~\cite{Paper-de-Broglie-5} Sobre os espectros corpusculares dos elementos (tamb\'{e}m com o irm\~{a}o Maurice)~\cite{Paper-de-Broglie-6}, Sobre a degrada\c{c}\~{a}o do \textit{quantum} nas transforma\c{c}\~{o}es sucessivas das radia\c{c}\~{o}es de alta frequ\^{e}ncia~\cite{Paper-de-Broglie-7}, Sobre a teoria de absor\c{c}\~{a}o dos raios X pela mat\'{e}ria e o princ\'{\i}pio de correspond\^{e}ncia~\cite{Paper-de-Broglie-8}, Sobre o sistema espectral dos raios de R\"{o}ntgen~\cite{Paper-de-Broglie-9}, Observa\c{c}\~{o}es sobre os espectros corpusculares e o efeito fotoel\'{e}trico (com seu irm\~{a}o)~\cite{Paper-de-Broglie-10}, \textit{Quanta} de luz, difra\c{c}\~{a}o e interfer\^{e}ncias~\cite{Paper-de-Broglie-11}, Os \textit{quanta}, a teoria cin\'{e}tica dos gases e o princ\'{\i}pio de Fermat~\cite{Paper-de-Broglie-12}.

\section{Alguma pistas abandonadas}
\paragraph*{}

De Broglie, ao ampliar o dom\'{\i}nio da crise, antecipa que, na microf\'{\i}sica, deve haver uma profunda rela\c{c}\~{a}o entre a descri\c{c}\~{a}o te\'{o}rica da radia\c{c}\~{a}o e da mat\'{e}ria, ou, mais especificamente, do f\'{o}ton e do el\'{e}tron, na qual cada uma destas part\'{\i}culas apresenta um car\'{a}ter dual, que extrapola suas descri\c{c}\~{o}es cl\'{a}ssicas. De um ponto de vista exclusivamente filos\'{o}fico, essa \textit{dualidade} espelha uma crise do que \'{e} o \textit{Ser}, definido, desde os prim\'{o}rdios da Filosofia Grega, sem ambiguidades. De fato, Planck e Einstein v\~{a}o mostrar que a luz, entidade f\'{\i}sica vista, ao final do s\'{e}culo XIX, como um fen\^{o}meno ondulat\'{o}rio, pode apresentar-se como um fen\^{o}meno corpuscular, como no efeito fotoel\'{e}trico.
\newpage

Esta breve Se\c{c}\~{a}o cumpre o papel de destacar tr\^{e}s resultados cl\'{a}ssicos que, de certa forma, podem ser vistos como pistas que sugerem uma poss\'{\i}vel compatibilidade para as duas descri\c{c}\~{o}es: o princ\'{\i}pio de superposi\c{c}\~{a}o dos fen\^{o}menos ondulat\'{o}rios estacion\'{a}rios; a deriva\c{c}\~{a}o das leis emp\'{\i}ricas de Stefan e Wien para a radia\c{c}\~{a}o de corpo negro, baseada no efeito Doppler e a descri\c{c}\~{a}o pr\'{e} relativ\'{\i}stica da Aberra\c{c}\~{a}o estelar.

Em uma corda homog\^{e}nea de densidade linear $\rho$ e comprimento $L$, sujeita a uma for\c{c}a de tens\~{a}o $F$, pode resultar um movimento ondulat\'{o}rio no qual todos os pontos da corda oscilem harmonicamente com mesma frequ\^{e}ncia ($\omega$) e constante de fase, tal que a fun\c{c}\~{a}o de onda possa ser expressa, por exemplo, como
\begin{equation}\label{sol_esta}
\Psi(x,t)= C\, \mbox{sen}\, kx \cos \omega t
\end{equation}
na qual $C$ \'{e} uma constante, e $\omega/k=v=(F/\rho)^{1/2}$ \'{e} a velocidade de propaga\c{c}\~{a}o de qualquer onda harm\^{o}nica que se propaga na corda (meio homog\^{e}neo e n\~{a}o dispersivo). Tal velocidade \'{e} denominada velocidade de fase.

A express\~{a}o anterior, equa\c{c}\~{a}o~(\ref{sol_esta}), satisfaz \`{a} equa\c{c}\~{a}o de onda de d'Alembert que descreve uma \textit{onda estacion\'{a}ria}, resultante da superposi\c{c}\~{a}o de duas ondas harm\^{o}nicas que se propagam em sentidos opostos com velocidades de magnitude $v$, que foram refletidas em seus extremos, supostamente fixos,
\begin{eqnarray*}
\Psi(x,t) &=& A \ \mbox{sen}\, (kx-\omega t+\phi_{_A})
+ B \cos\, (kx + \omega t+\phi_{_B}) \\
&=& a(t)\ \mbox{sen}\, kx \ + \ b(t) \cos kx
\end{eqnarray*}
\noindent em que $A$ e $B$ s\~{a}o constantes e $a(t)$ e $b(t)$ s\~{a}o combina\c{c}\~{o}es lineares de $\cos \omega t$ e $\ \mbox{sen} \, \omega t$. Agora, quando se toma um dos extremos como origem, se imp\~{o}em as condi\c{c}\~{o}es de contorno
$$
\mbox{(extremos fixos)} \qquad
\left\{ \begin{array}{l}
\Psi(0,t) = 0 \quad \Longrightarrow \quad b(t)=0  \\
\ \\
\Psi(L,t) = 0
\end{array}
\right.
$$
e as condi\c{c}\~{o}es iniciais
$$\left\{ \begin{array}{l}
\Psi(x,0) = C \ \mbox{sen}\,kx \qquad \mbox{(perfil inicial)}
\quad \Longrightarrow \quad a(0)=C \\
\ \\
\displaystyle
\frac{\partial \Psi}{\partial t}(x,0) = 0
\qquad \mbox{(velocidade inicial)}
\quad \Longrightarrow \quad a(t)=C\cos \omega t
\vspace{.2cm}
\end{array}
\right.
$$
e, assim, obt\'{e}m-se a equa\c{c}\~{a}o~(\ref{sol_esta}).

Embora a velocidade de fase das componentes harm\^{o}nicas n\~{a}o dependa das condi\c{c}\~{o}es de con\-tor\-no, o mesmo n\~{a}o ocorre para o comprimento de onda e a frequ\^{e}ncia, porque a condi\c{c}\~{a}o para o extremo $x=L$ implica que
$$ \mbox{sen}\, kL = 0 \quad \Longrightarrow \quad kL = n \pi \quad (n=1,2,3,\ldots) $$
ou seja,
$$ \lambda_n = \frac{2\pi}{k} = \frac{2L}{n} \qquad \mbox{e} \qquad \nu_n = \frac{v}{\lambda_n} = \frac{n}{2L} \sqrt{\frac{F}{\rho}} $$

\index{modos! normais de vibra\c{c}\~{a}o}
Nesse caso, os valores poss\'{\i}veis para o comprimento de onda e a frequ\^{e}ncia constituem um conjunto nu\-m\'{e}\-rico discreto, de comprimentos de onda e frequ\^{e}ncias pr\'{o}prios, ou caracter\'{\i}sticos,
$$\left\{ \lambda_n, \nu_n \right\}\qquad (n=1,2,3,\ldots)$$
que se relacionam a um conjunto de  solu\c{c}\~{o}es estacion\'{a}rias $\Psi_n(x,t)$, denominadas \textit{modos normais de vi\-bra\-\c{c}\~{a}o},\footnote{\, Tamb\'{e}m denominadas \textit{fun\c{c}\~{o}es pr\'{o}prias} ou \textit{autofun\c{c}\~{o}es}, $\Psi_n(x,t)$, da equa\c{c}\~{a}o de onda de d'Alembert, que satisfazem as condi\c{c}\~{o}es de contorno homog\^{e}neas em dois pontos $0$ e $L$, isto \'{e}, $\Psi_n(0,t)= \Psi_n(L,t) = 0 $.} dadas por
\index{equa\c{c}\~{a}o! de onda}
\index{equa\c{c}\~{a}o! de onda! de d'Alembert}
\index{d'Alembert! equa\c{c}\~{a}o de onda de}
$$ \Psi_n(x,t) = C_n \ \mbox{sen} \left( \frac{n\pi}{L} x\right) \cos \omega_n t $$
sendo $\omega_n = 2\pi \nu_n$.

\index{equa\c{c}\~{a}o! de onda}
\index{equa\c{c}\~{a}o! de onda! de d'Alembert}
\index{d'Alembert! equa\c{c}\~{a}o de onda de}
Devido \`{a} linearidade da equa\c{c}\~{a}o de onda de d'Alembert, a solu\c{c}\~{a}o mais geral para as ondas es\-ta\-cion\'{a}rias \'{e} dada pela superposi\c{c}\~{a}o linear dos modos normais
\begin{equation}\label{eq_corda_vib}
\Psi(x,t) = \sum_{n=1}^\infty \Psi_n(x,t) = \sum_{n=1}^\infty C_n \
\mbox{sen} \left( \frac{n\pi}{L} x\right) \cos \omega_n t
\end{equation}

No instante $t=0$, $\Psi(x,t)$ se reduz ao perfil inicial, $f(x)$, da corda, ou seja,
\begin{equation}\label{perfil_inic}
\Psi(x,0) = \displaystyle f(x) = \sum_{n=1}^\infty C_n \, \psi_n(x)
\end{equation}
em que $\displaystyle \psi_n (x) = \mbox{sen}\ (n\pi x/L)$.

\index{nome Euler (1707-1783)! Leonhard}
\index{nome Lagrange (1736-1813)! Joseph Louis}
\index{nome Bernoulli (1700-1782)! Daniel}
\index{nome Fourier (1768-1830)! Jean-Baptiste Joseph}
Uma vez que o perfil inicial da corda \'{e} arbitr\'{a}rio, a express\~{a}o (\ref{perfil_inic}) sugere que qualquer fun\c{c}\~{a}o \'{e} capaz de ser descrita, em um dado intervalo, por uma s\'{e}rie de fun\c{c}\~{o}es harm\^{o}nicas. Essa marcante possibilidade, j\'{a} utilizada por D.~Bernoulli, Euler e Lagrange, foi amplamente utilizada por Jean-Baptiste Joseph Fourier, em 1807, em sua obra sobre a propaga\c{c}\~{a}o do calor. A representa\c{c}\~{a}o de uma fun\c{c}\~{a}o arbitr\'{a}ria por s\'{e}ries de fun\c{c}\~{o}es harm\^{o}nicas \'{e} conhecida como \textit{s\'{e}rie de Fourier}.
\index{nome Riemann (1826-1866)! Bernhard}
\index{nome Lebesgue (1875-1941)! Henri}

Pode-se mostrar que a energia total armazenada na corda vibrante ($\epsilon$), igual \`{a} energia cin\'{e}tica ($\epsilon_c$) da corda mais a sua energia potencial ($\epsilon_p$) \'{e} dada por~\cite{Caruso-Oguri}:
\begin{equation}\label{E-sum-e-n}
\epsilon =\epsilon_c + \epsilon_p = \sum_{n=1}^{\infty} \frac{1}{4} (\rho L) \ \omega_n^2 \ C_n^2 \, \equiv \sum_{n=1}^{\infty} \epsilon_n
\end{equation}
Essa igualdade revela o fato de que a energia total \'{e} a soma das energias dos v\'{a}rios modos normais de vibra\c{c}\~{o}es da corda, ou seja, estes modos normais comportam-se como componentes independentes pelos quais a energia total se acha distribu\'{\i}da.

A equa\c{c}\~{a}o~(\ref{E-sum-e-n}) pode ser interpretada, portanto, considerando-se a estrutura discreta de uma corda el\'{a}stica. Em outras palavras, \'{e} poss\'{\i}vel descrever o fen\^{o}meno da propaga\c{c}\~{a}o de ondas ac\'{u}sticas pelas vibra\c{c}\~{o}es de suas part\'{\i}culas constituintes, tal que a energia total resultaria da soma das energias de um conjunto de osciladores independentes, cada qual com frequ\^{e}ncia pr\'{o}pria $\nu$. Como se sabe da Mec\^{a}nica Qu\^{a}ntica, a menos de uma constante, a energia de cada oscilador ($\epsilon_n$) \'{e} um m\'{u}ltiplo de sua frequ\^{e}ncia pr\'{o}pria ($\epsilon_n \propto n \nu$), sendo $h$ de Planck a constante de proporcionalidade. Assim, pode-se considerar que cada estado din\^{a}mico de um oscilador, associado a uma energia $\epsilon_n$, corresponde a $n$ ``part\'{\i}culas'', cada qual com energia $h\nu$. O comportamento din\^{a}mico dessas ``part\'{\i}culas'', denominadas \textit{f\^{o}nons}, descreve as propriedades dos fen\^{o}menos ondulat\'{o}rios em um meio el\'{a}stico como a corda vibrante~\cite{Caruso-Oguri}.

A segunda pista refere-se a como, em 1893, o f\'{\i}sico alem\~{a}o Wilhelm Wien chegou \`{a} sua famosa express\~{a}o para a densidade de energia, $u_\lambda$, da radia\c{c}\~{a}o de um corpo negro
\begin{equation}\label{Wien_law}
\displaystyle \ \ u_{\nu} = \nu^3 \phi\left(\frac{\nu}{T}\right)
\end{equation}
em que $\phi$ \'{e} uma fun\c{c}\~{a}o da raz\~{a}o entre a frequ\^{e}ncia $\nu$ da radia\c{c}\~{a}o e a temperatura $T$ do corpo negro.

Admitindo, como se faz hoje, que um corpo negro seja modelado por uma esfera oca, de raio $r$, cujas paredes s\~{a}o perfeitamente condutoras contendo um pequeno orif\'{\i}cio, Wien, para chegar \`{a} express\~{a}o~(\ref{Wien_law}), presumiu correta a lei de Stefan e desenvolveu um engenhoso argumento, fundamentado na invari\^{a}ncia de escala, no qual admite que as paredes do corpo negro possam se contrair. Com base nisso, \'{e} poss\'{\i}vel determinar, a partir de um longo c\'{a}lculo, envolvendo o efeito Doppler da reflex\~{a}o da parede m\'{o}vel sobre a radia\c{c}\~{a}o, n\~{a}o reproduzido aqui~\cite{Caruso-Oguri}, como o comprimento de onda $\lambda$ e a frequ\^{e}ncia $\nu$ da radia\c{c}\~{a}o varia com a varia\c{c}\~{a}o do raio $r(t)$. O ponto que interessa para nosso argumento \'{e} que a f\'{o}rmula de Wien s\'{o} \'{e} obtida dessa forma com a hip\'{o}tese de que a varia\c{c}\~{a}o percentual da energia da radia\c{c}\~{a}o, para um dado comprimento de onda, \'{e} a mesma que a varia\c{c}\~{a}o percentual da frequ\^{e}ncia, ou seja,
\begin{equation}\label{delta_epsilon_relat2}
\frac{\delta\epsilon}{\epsilon}= \frac{\delta\nu}{\nu}
\end{equation}

Isso significa que a rela\c{c}\~{a}o entre as grandezas $\epsilon$ e $\nu$ \'{e} \textit{linear}, mesmo quando a luz \'{e} con\-si\-derada uma onda. Portanto, verifica-se, j\'{a} nesse ponto, que n\~{a}o h\'{a} contradi\c{c}\~{a}o, no que se refere \`{a} luz, entre o resultado cl\'{a}ssico e a proposta posterior de Planck, visto que a rela\c{c}\~{a}o linear entre essas duas grandezas \'{e} mantida, com a introdu\c{c}\~{a}o de uma nova constante $h$, isto \'{e}, $\epsilon = h \nu$.

A terceira e \'{u}ltima pista, sem a preocupa\c{c}\~{a}o com a ordem cronol\'{o}gica dos fatos, tem origem na explica\c{c}\~{a}o te\'{o}rica da Aberra\c{c}\~{a}o estelar, descoberto, em 1728, pelo astr\^{o}nomo ingl\^{e}s James Bradley~\cite{Bradley}. Sua explica\c{c}\~{a}o original, devida ao pr\'{o}prio Bradley, considerava ainda o car\'{a}ter corpuscular da luz. No entanto, posteriormente, os partid\'{a}rios da vis\~{a}o ondulat\'{o}ria da luz, lan\c{c}ando m\~{a}o de diferentes concep\c{c}\~{o}es de \'{e}ter, explicavam igualmente este fen\^{o}meno. Tanto a imagem corpuscular quanto ondulat\'{o}ria da luz, uma vez mais, levam a um mesmo resultado. Esse ponto s\'{o} foi elucidado com a Teoria da Relatividade Especial de Einstein, em 1905.

\section{As nuvens de Lorde Kelvin: o in\'{\i}cio da crise}
\paragraph*{}
No clima de virada do s\'{e}culo~XIX para o XX, em uma confer\^{e}ncia proferida em 27 de abril de 1900, Lorde Kelvin, partid\'{a}rio da vis\~{a}o mecanicista, afirmou que, no c\'{e}u azul da F\'{\i}sica Cl\'{a}ssica, existiam duas nuvens: o problema da n\~{a}o detec\c{c}\~{a}o do vento de \textit{\'{e}ter} e o problema da parti\c{c}\~{a}o de energia. Seu artigo, publicado em 1901, dedicado \`{a}s ``nuvens'' do s\'{e}culo~XIX, se inicia com essa coloca\c{c}\~{a}o:
\index{nome Young (1773-1829)! Thomas}
\index{nome Fresnel (1788-1821)! Augustin-Jean}
\begin{quotation}
\noindent \baselineskip=10pt {\small
\textit{A beleza e a clareza da teoria din\^{a}mica, que sustenta que calor e luz s\~{a}o modos de movimento, no presente \'{e} obscurecida {\em [sic.]} por duas nuvens. I. A primeira relaciona-se com a teoria ondulat\'{o}ria da luz, e foi tratada por Fresnel e pelo Dr.~Thomas Young; ela envolve a quest\~{a}o `Como pode a Terra mover-se atrav\'{e}s de um s\'{o}lido el\'{a}stico, tal como \'{e} essencialmente o \'{e}ter lumin\'{\i}fero?' II. A segunda \'{e} a doutrina de Maxwell-Boltzmann, referente \`{a} parti\c{c}\~{a}o de energia.}}
\end{quotation}
\index{nome Maxwell (1831-1875)! James Clerk}
\index{nome Boltzmann (1844-1906)! Ludwig}

\index{nome Kelvin (1824-1907)! Lorde (William Thomson)}
Esse \'{e} um exemplo not\'{a}vel, no qual a presun\c{c}\~{a}o se mistura \`{a} perspic\'{a}cia. Decerto, estimulado pelo es\-p\'{\i}\-rito de que o fim de um s\'{e}culo marca o \textit{fim} de muita coisa, Lorde Kelvin, por um lado, apontou que o conjunto de teorias que se convencionou chamar de F\'{\i}sica Cl\'{a}ssica dava conta de praticamente \textit{todos} os fen\^{o}menos observados e, por outro lado, foi capaz de identificar exatamente os \textit{dois} desafios mais significativos de sua \'{e}poca, embri\~{o}es, sabe-se hoje, de duas grandes revolu\c{c}\~{o}es cient\'{\i}ficas.

A dissipa\c{c}\~{a}o dessas duas nuvens foi o ponto de partida de uma renova\c{c}\~{a}o radical de conceitos na F\'{\i}sica, que resultou na constru\c{c}\~{a}o e cria\c{c}\~{a}o das \textit{teorias qu\^{a}nticas e relativ\'{\i}sticas}. Conceitos e defini\c{c}\~{o}es como os de \textit{espa\c{c}o}, \textit{tempo}, \textit{simultaneidade}, \textit{energia}, \textit{massa}, \textit{trajet\'{o}ria}, \textit{part\'{\i}cula}, \textit{intera\c{c}\~{a}o} e \textit{vazio} foram revistas \`{a} luz dessas novas teorias.

\section{Resumo da contribui\c{c}\~{a}o de Planck e de Einstein \`{a} nova F\'{\i}\-sica}
\paragraph*{}

A an\'{a}lise espectrosc\'{o}pica da emiss\~{a}o da radia\c{c}\~{a}o de corpo negro, empreendida pelo f\'{\i}sico alem\~{a}o Friedrich Paschen, em 1894, envolvia comprimentos de onda relativamente curtos, de $\simeq 5$ $\mu$m, na faixa do infravermelho. De suas observa\c{c}\~{o}es, Paschen e Wien sugeriram, independentemente, em 1896, uma f\'{o}rmula semiemp\'{\i}rica que se ajustava \`{a}s curvas experimentais da intensidade da radia\c{c}\~{a}o emitida.

Apesar do sucesso inicial da f\'{o}rmula de Wien, suas limita\c{c}\~{o}es foram rapidamente evidenciadas quando, no in\'{\i}cio do s\'{e}culo~XX, mais especificamente em 1900, os dois gru\-pos do \textit{Physicalish-Technische Reichsanstalt}, de Berlim, constitu\'{\i}dos de Otto Lummer, Ernst Pringsheim, Ferdinand Kurlbaum e Heinrich Rubens, estenderam as observa\c{c}\~{o}es para comprimentos de onda maiores, inicialmente at\'{e} 18~$\mu$m e, logo ap\'{o}s, na faixa de 30 a 50 $\mu$m, a temperaturas entre 200$^{\mbox{\scriptsize o}}$C e 1\,600$^{\mbox{\scriptsize o}}$C. Os resultados obtidos dessa maneira, principalmente por Kurlbaum e Rubens, estabeleceram definitivamente que, para essas frequ\^{e}ncias menores, bem afastadas da regi\~{a}o vis\'{\i}vel, em vez da f\'{o}rmula de Wien, a rec\'{e}m-proposta f\'{o}rmula de Rayleigh era a que mais adequadamente se ajustava aos dados.

Cabe destacar que, apesar de compat\'{\i}vel com os resultados emp\'{\i}ricos de Rubens e Kurlbaum, no dom\'{\i}nio de longos comprimentos de onda (frequ\^{e}ncias baixas), a express\~{a}o de Rayleigh implica o limite
\[
\lim_{\nu\rightarrow \infty} \ u_{\nu} (T) = \infty
\]

O fato de que, para altas frequ\^{e}ncias, ou seja, para pequenos comprimentos de onda, a densidade de energia da radia\c{c}\~{a}o prevista pela f\'{o}rmula de Rayleigh fosse bem maior que a obtida experimentalmente, tendendo mesmo a valores infinitos, ficou conhecido, alguns anos mais tarde, como ``a cat\'{a}strofe do ultravioleta'', express\~{a}o cunhada pelo f\'{\i}sico austr\'{\i}aco Paul Ehrenfest, em 1911. Essa denomina\c{c}\~{a}o reflete o espanto gerado pelo insucesso da abordagem cl\'{a}ssica de Rayleigh-Jeans ao problema.

Cabe destacar que, ao deduzir a f\'{o}rmula da radia\c{c}\~{a}o de corpo negro, Planck introduziu duas constantes universais, a constante de Planck ($h$) e a constante de Boltzmann ($k$), que podem ser determinadas a partir das leis de deslocamento ($\lambda_M T = b$) e de Stefan ($I = \sigma T^4$), e apresentou tamb\'{e}m estimativas para o n\'{u}mero de Avogadro e para a carga elementar, com base nas das rela\c{c}\~{o}es $N_A = R/k$ e $e = F/N_A$, sendo $R \simeq 8,\!3$ J/K e $F \simeq 96\,500$ C/mol.

\begin{figure}[htbp]
\centerline{\includegraphics[width=9.5cm]{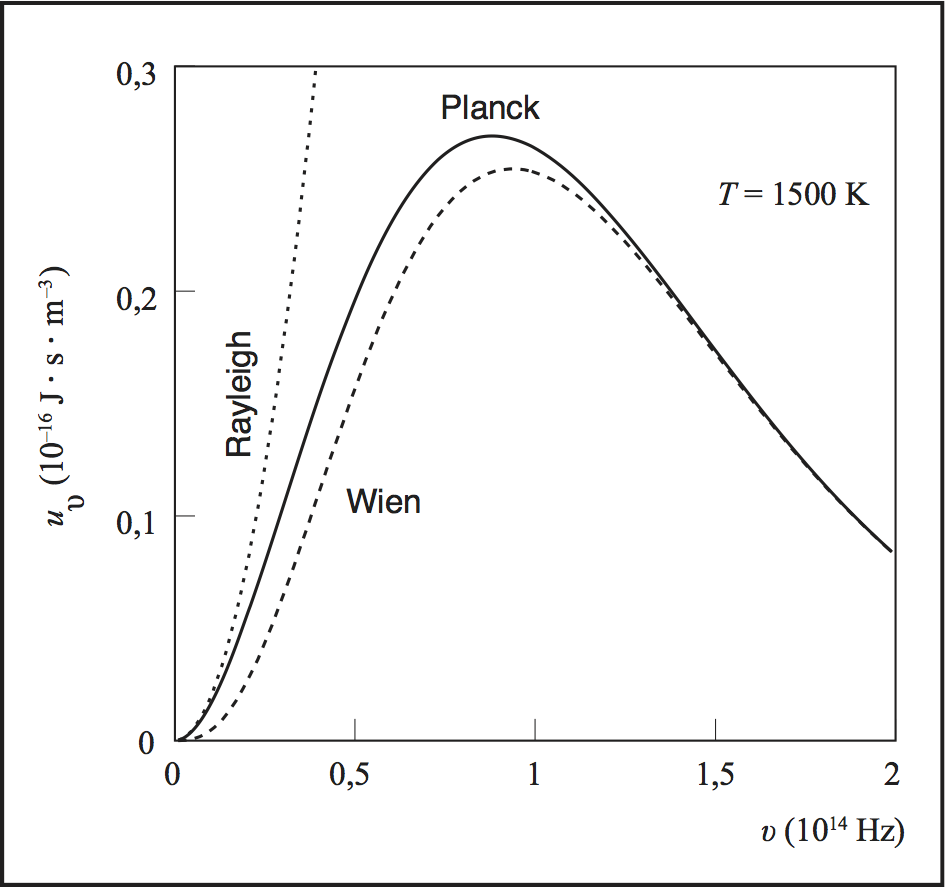}}
\caption{\baselineskip=8pt{\small Compara\c{c}\~{a}o da lei de Planck com as predi\c{c}\~{o}es das f\'{o}rmulas de Rayleigh e Wien.}}
\label{comparacao}
\end{figure}

Com rela\c{c}\~{a}o \`{a} constante de Planck, ele pr\'{o}prio se refere assim: \textit{Tentei imediatamente in\-corporar de alguma forma o {\em quantum} elementar de a\c{c}\~{a}o `h' no contexto da teoria cl\'{a}s\-si\-ca. Mas, em face de todas essas tentativas, esta constante mostrou-se obstinada.}

J\'{a} o f\'{\i}sico israelense e historiador da F\'{\i}sica Max Jammer sintetiza o impacto da con\-tri\-bui\c{c}\~{a}o de Planck com as seguintes palavras: \textit{Nunca na Hist\'{o}ria da F\'{\i}sica houve uma in\-ter\-pola\c{c}\~{a}o matem\'{a}tica t\~{a}o impercept\'{\i}vel com t\~{a}o amplas consequ\^{e}ncias f\'{\i}sicas e filos\'{o}ficas.}

A conhecida f\'{o}rmula para a densidade espectral de energia de um corpo negro, que interpola os limites de Rayleigh e de Wien, obtida por Planck, \'{e}
\begin{equation} \label{u_Planck}
\fbox{
$\displaystyle  \ \ u_{\nu} = \frac{8\pi h\nu^3}{c^3} \
\frac{1}{e^{h \nu /kT}-1} \ \ $}
\end{equation}
ou, em termos do comprimento de onda,

\begin{equation}\label{ul_Planck}
\fbox{
$\displaystyle \ \ u_{\lambda} = \frac{8\pi c h}{\lambda^5} \
\frac{1}{e^{hc/k\lambda T}-1} \ \ $}
\end{equation}
e pressup\~{o}e a validade do postulado
\vspace*{0.5cm}

\centerline{
\fbox{
$\displaystyle  \ \
\epsilon_n = n \epsilon_{_0} = n h\nu \qquad  \qquad (n = 0,1,2,\cdots)
\ \ $}}

Para Planck, a circunst\^{a}ncia de que a energia de um oscilador em equil\'{\i}brio t\'{e}rmico com a radia\c{c}\~{a}o s\'{o} pudesse ser trocada em quantidades discretas, m\'{u}ltiplas de um \textit{quantum} de energia, era um efeito que s\'{o} se manifestaria na intera\c{c}\~{a}o de ondas eletromagn\'{e}ticas, confinadas em uma regi\~{a}o, com a mat\'{e}ria. Quem realmente de\-fen\-deu a conjectura da quantiza\c{c}\~{a}o da energia de um oscilador, independentemente de sua intera\c{c}\~{a}o com a radia\c{c}\~{a}o, durante a primeira d\'{e}cada do s\'{e}culo~XX, foi Einstein, em 1907, ao explicar o comportamento dos calores espec\'{\i}ficos dos s\'{o}lidos. Antes, por\'{e}m, em um dos artigos de 1905, pelo qual se costuma dizer que Einstein foi agraciado com o Pr\^{e}mio Nobel, o efeito fotoel\'{e}trico foi explicado em termos de f\'{o}tons. Mas mais do que isso, talvez, um resultado igualmente importante foi ter sido mostrado que a luz no interior de uma cavidade refletora (modelo de um corpo negro) pode ser descrita como um g\'{a}s de f\'{o}tons. Portanto, o f\'{\i}sico alem\~{a}o n\~{a}o era avesso \`{a} utiliza\c{c}\~{a}o da estat\'{\i}stica na F\'{\i}sica, desde que o sistema constasse de um n\'{u}mero enorme de constituintes.

Admitindo-se a hip\'{o}tese de quantiza\c{c}\~{a}o da luz de Einstein, pode-se compreender, qualitativamente, os espectros discretos de emiss\~{a}o e de absor\c{c}\~{a}o dos gases. Considerando que o fen\^{o}meno resulta de v\'{a}rios processos discretos, envolvendo a troca de energia entre um \'{a}tomo e um f\'{o}ton, a energia do \'{a}tomo ap\'{o}s a emiss\~{a}o ($\epsilon^\prime$) deve ser menor do que aquela ($\epsilon$) antes, de tal forma que a diferen\c{c}a ($\epsilon - \epsilon'$) seja igual \`{a} energia ($\epsilon_\gamma$) do f\'{o}ton emitido.

Uma vez que a energia de um f\'{o}ton \'{e} proporcional \`{a} frequ\^{e}ncia da radia\c{c}\~{a}o emitida, isso implica
$$ |\epsilon - \epsilon^\prime| \sim \nu $$

O espectro de linhas de um g\'{a}s, determinado pelas frequ\^{e}ncias caracter\'{\i}sticas de cada subst\^{a}ncia, implica, por sua vez, que as energias dos \'{a}tomos dessas subst\^{a}ncias constituam igualmente um conjunto discreto de valores, ou seja, conclui-se, da espectroscopia, que os \'{a}tomos dos gases possuem um espectro de energia discreto.

Coube a Bohr, em 1913, introduzir um modelo din\^{a}mico para o movimento do el\'{e}tron em um \'{a}tomo de hidrog\^{e}nio, capaz de prever as frequ\^{e}ncias caracter\'{\i}sticas daquele \'{a}tomo, ou os n\'{\i}veis de energia associados ao seu espectro (Se\c{c}\~{a}o~\ref{Bohr}).

\section{Bohr e a estabilidade do \'{a}tomo: primeira amplia\c{c}\~{a}o da crise}\label{Bohr}
\paragraph*{}

De maneira an\'{a}loga ao que fez Planck, Bohr concentrou-se no emissor da radia\c{c}\~{a}o, o \'{a}tomo, e, assim, con\-se\-guiu explicar por que apenas certas linhas s\~{a}o exibidas no espectro do hidrog\^{e}nio, dando uma coe\-r\^{e}ncia te\'{o}rica \`{a} f\'{o}rmula de Balmer.

O f\'{\i}sico dinamarqu\^{e}s Niels Bohr, em 1913, estabeleceu que o \textit{quantum} elementar de a\c{c}\~{a}o, expresso pela constante de Planck, ori\-ginalmente introduzido para explicar a emiss\~{a}o de luz de um corpo negro, tamb\'{e}m seria necess\'{a}rio para garantir a \textit{estabilidade da mat\'{e}ria}. Seu modelo at\^{o}mico foi, provavelmente, uma das concep\c{c}\~{o}es mais intrigantes e frut\'{\i}feras acerca da estrutura interna dos elementos qu\'{\i}micos. Essa opini\~{a}o foi corroborada por Dirac, quem declarou sua convic\c{c}\~{a}o de que \textit{a introdu\c{c}\~{a}o dessas ideias por Bohr foi o maior passo de todos no desenvolvimento da Mec\^{a}nica Qu\^{a}ntica.}

A solu\c{c}\~{a}o da instabilidade din\^{a}mica do \'{a}tomo de Rutherford, proposta por Bohr, estava em contradi\c{c}\~{a}o direta com o que se esperava da teoria do movimento das cargas aceleradas de Maxwell-Lorentz, ampliando a crise da F\'{\i}sica Cl\'{a}ssica (estendendo-a \`{a} mat\'{e}ria), que havia sido iniciada, alguns anos antes, pelos dois artigos seminais de Planck e Einstein, os quais lan\c{c}aram luz sobre ambas as nuvens de Lorde Kelvin. Nas pr\'{o}prias palavras de Bohr, isso \textit{parece ser um reconhecimento geral da inadequa\c{c}\~{a}o da eletrodin\^{a}mica cl\'{a}ssica em descrever o comportamento de sistemas de tamanho at\^{o}mico.}

Entrementes, o modelo de Bohr foi muito bem sucedido em reproduzir a regularidade dos espectros at\^{o}\-micos, especialmente as regularidades do espectro de Balmer para \'{a}tomos de hidrog\^{e}nio e para os hidrogen\'{o}ides, como os vapores de metais alcalinos.\footnote{\, S\~{a}o os elementos qu\'{\i}micos do grupo 1 da Tabela Peri\'{o}dica: l\'{\i}tio ({\tt Li}), s\'{o}dio ({\tt Na}), pot\'{a}ssio ({\tt K}), rub\'{\i}dio ({\tt Rb}) e c\'{e}sio ({\tt Cs}), excetuando-se o hidrog\^{e}nio ({\tt H}), e possuem um \'{u}nico el\'{e}tron de val\^{e}ncia.} Referindo-se a esse sucesso, at\'{e} ent\~{a}o n\~{a}o compreens\'{\i}vel, James Jeans foi o primeiro a reconhecer publicamente o valor e a engenhosidade do trabalho do f\'{\i}sico dinamarqu\^{e}s.

Niels Bohr foi quem primeiro percebeu que a estabilidade da mat\'{e}ria dependeria da introdu\c{c}\~{a}o da constante de Planck na descri\c{c}\~{a}o do \'{a}tomo, ou seja, era preciso formular um modelo qu\^{a}ntico para o \'{a}tomo, pois o modelo de Rutherford era mecanicamente inst\'{a}vel. Concentrado na emiss\~{a}o de luz pelo \'{a}tomo de hidrog\^{e}nio, sua aten\c{c}\~{a}o se volta para o processo elementar envolvido. Segundo Edmund Whittaker~\cite{Whittaker}, essas s\~{a}o as hip\'{o}teses de Bohr:
\begin{itemize}
  \item os \'{a}tomos produzem as linhas espectrais uma de cada vez;
  \item o \'{a}tomo de Rutherford oferece uma base satisfat\'{o}ria para os c\'{a}lculos exatos dos comprimentos de onda das linhas espectrais;
  \item a produ\c{c}\~{a}o dos espectros at\^{o}micos \'{e} um fen\^{o}meno qu\^{a}ntico;
  \item um simples el\'{e}tron \'{e} o agente desse processo;
  \item dois estados distintos do \'{a}tomo est\~{a}o envolvidos na produ\c{c}\~{a}o de uma linha espectral;
  \item a rela\c{c}\~{a}o $E = h\nu$, correlacionando a energia e a frequ\^{e}ncia da radia\c{c}\~{a}o, \'{e} v\'{a}lida tanto para a emiss\~{a}o como para a absor\c{c}\~{a}o.
\end{itemize}

Dessa forma, ele contribui para a generaliza\c{c}\~{a}o da crise da F\'{\i}sica Cl\'{a}ssica contida na express\~{a}o \textit{dualidade onda-part\'{\i}cula}. Louis de Broglie vai associar uma onda ao el\'{e}tron cujo comprimento de onda ser\'{a} $\lambda = h/p$. Portanto, se antes s\'{o} a luz poderia, de acordo com o tipo de experimento, apresentar um car\'{a}ter dual, agora tamb\'{e}m o el\'{e}tron dever\'{a} se comportar ora como onda, ora como corp\'{u}sculo, o que foi efetivamente observado \cite{Davisson1,Davisson2}.\footnote{\, Essa dualidade se apresenta quando o el\'{e}tron e o f\'{o}ton s\~{a}o descritos de modo cl\'{a}ssico ou h\'{\i}brido, ou seja, por teorias semicl\'{a}ssicas com argumentos \textit{ad hoc}.}

Por outro lado, os postulados de Bohr pressup\~{o}em uma renuncia a todas as tentativas de visualizar ou de explicar classicamente o comportamento do el\'{e}tron agente do processo durante uma transi\c{c}\~{a}o do \'{a}tomo entre um estado estacion\'{a}rio e outro. Segundo o testemunho de Dirac~\cite{Directions},
\begin{quotation}
\noindent\baselineskip=11pt{\small{Heisenberg disse que as \'{o}rbitas de Bohr n\~{a}o s\~{a}o muito importantes. As coisas que s\~{a}o observadas, ou que est\~{a}o estreitamente conectadas com as quantidades observadas, est\~{a}o todas associadas com duas \'{o}rbitas de Bohr e n\~{a}o apenas com uma \'{o}rbita de Bohr: \textit{duas e n\~{a}o uma}.\footnote{O grifo \'{e} nosso.}}}
\end{quotation}

Vemos aqui a g\^{e}nese da ideia de que a F\'{\i}sica Qu\^{a}ntica deve tratar apenas dos observ\'{a}veis, segundo o ponto de vista de Heisenberg, que n\~{a}o nos parece ser estranho ao pensamento que norteou Bohr, doze anos antes, a propor o primeiro modelo qu\^{a}ntico do \'{a}tomo. O observ\'{a}vel -- o espectro de Balmer, no caso -- depende da emiss\~{a}o de f\'{o}tons pela transi\c{c}\~{a}o do el\'{e}tron entre \textit{duas} \'{o}rbitas. Apenas isso \'{e} observ\'{a}vel; a \'{o}rbita em si, n\~{a}o.

\section{Louis de Broglie: a crise dentro da crise}\label{de_Broglie}
\paragraph*{}

Tanto Louis de Broglie quanto Erwin Schr\"{o}dinger deram contribui\c{c}\~{o}es \`{a} compreens\~{a}o da F\'{\i}sica Qu\^{a}ntica baseados nos estudos de Hamilton, de 1835, nos quais estabelecem-se analogias formais entre a \'{O}ptica e a Mec\^{a}nica Cl\'{a}ssica. Nas palavras do pr\'{o}prio de Broglie~\cite{Caruso-Oguri},

\begin{quotation}
\noindent\baselineskip=10pt {\small
[\textit{Schr\"{o}dinger}] \textit{aprofundando a analogia assinalada\, {\rm [...]} por Hamilton, entre a \'{O}ptica Geo\-m\'{e}trica e a Mec\^{a}nica Anal\'{\i}tica, conseguiu escrever a equa\c{c}\~{a}o geral de propaga\c{c}\~{a}o, v\'{a}lida na aproxima\c{c}\~{a}o n\~{a}o relativ\'{\i}stica, para uma onda associada a um corp\'{u}sculo em um dado campo {\rm [...]}.}\footnote{\,Hamilton expressou a equa\c{c}\~{a}o de movimento de uma part\'{\i}cula de massa $m$, sob a a\c{c}\~{a}o de um campo de for\c{c}as, de um modo bastante similar \`{a}s equa\c{c}\~{o}es que descrevem a trajet\'{o}ria de um raio de luz em um meio n\~{a}o homog\^{e}neo, cujo \'{\i}ndice de refra\c{c}\~{a}o depende da posi\c{c}\~{a}o. As varia\c{c}\~{o}es do \'{\i}ndice de refra\c{c}\~{a}o modificam a trajet\'{o}ria dos raios luminosos da mesma maneira que a varia\c{c}\~{a}o da energia potencial de intera\c{c}\~{a}o faz com que as trajet\'{o}rias de part\'{\i}culas sejam curvas.
}}
\end{quotation}

\begin{figure}[hbtp]
\centerline{\includegraphics[width=7.0cm]{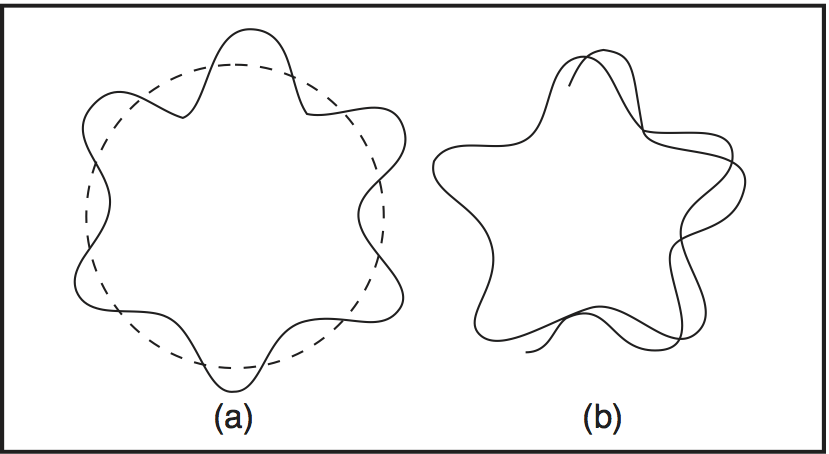}}
\caption{\baselineskip=6pt
{\small \textbf{\'{O}rbitas e ondas estacion\'{a}rias associadas ao movimento do el\'{e}tron em um \'{a}tomo.}}}
\label{onda_estacionaria_s}
\end{figure}

Para de Broglie, um el\'{e}tron ligado ao n\'{u}cleo de um \'{a}tomo descrevendo \'{o}rbitas circulares somente poderia ter associada a ele uma onda-piloto se esta fosse estacion\'{a}ria, o que preservaria a estabilidade do \'{a}tomo (Figura~\ref{onda_estacionaria_s}a), ao contr\'{a}rio do que ocorreria na situa\c{c}\~{a}o representada na Figura~\ref{onda_estacionaria_s}b, na qual haveria interfer\^{e}ncia destrutiva.

Nessas circunst\^{a}ncias, o per\'{\i}metro de uma poss\'{\i}vel \'{o}rbita deveria ser um m\'{u}ltiplo inteiro de com\-pri\-mentos de onda da onda-piloto, ou seja,
\[
\oint \ \frac{\mbox{d}s}{\lambda} = n \qquad \qquad (n=1,2,\ldots)
\]
na qual $\mbox{d}s$ \'{e} um elemento de arco ao longo da trajet\'{o}ria do el\'{e}tron.

Levando-se em conta a rela\c{c}\~{a}o entre o \textit{momentum} e o comprimento de onda, $p=h/\lambda$, obt\'{e}m-se
\[
\oint p \, \mbox{d}s = nh
\]
que \'{e}, essencialmente, a regra de quantiza\c{c}\~{a}o de Wilson-Sommerfeld. A rela\c{c}\~{a}o $p = \hbar k = h/\lambda$, que Einstein havia mostrado se aplicar aos f\'{o}tons, deveria, portanto, segundo de Broglie, ser igualmente v\'{a}lida para el\'{e}trons massivos.

Lembre-se que, alicer\c{c}ado nos trabalhos de Planck sobre a radia\c{c}\~{a}o de corpo negro, nos quais \'{e} introduzida a constante fundamental $h$, Einstein havia mostrado ser poss\'{\i}vel associar a uma onda eletromagn\'{e}tica plana monocrom\'{a}tica, de frequ\^{e}ncia $\nu$, um conjunto de part\'{\i}culas, os f\'{o}tons, que carregam, cada um, um fragmento ou \textit{quantum} de energia $E$, proporcional \`{a} frequ\^{e}ncia da radia\c{c}\~{a}o ($E = h \nu$), tal que a energia total da onda, em uma dada regi\~{a}o do espa\c{c}o, \'{e} expressa como a soma das energias dos f\'{o}tons. Nesse sentido, uma onda eletromagn\'{e}tica apresentaria uma natureza discreta, sendo constitu\'{\i}da de corp\'{u}sculos n\~{a}o materiais de energia: os \textit{f\'{o}tons}.

De maneira an\'{a}loga, L.~de Broglie considerou que, assim como a um conjunto de f\'{o}tons de energia $E$ associa-se uma onda eletromagn\'{e}tica de frequ\^{e}ncia $\nu = E/ h$, pode-se associar a um feixe de part\'{\i}culas livres de massa $m$, todas com a mesma velocidade, um comportamento ondulat\'{o}rio (uma onda de mat\'{e}ria).

Para concluir essa Se\c{c}\~{a}o, um coment\'{a}rio sobre a coer\^{e}ncia das vis\~{o}es corpuscular e ondulat\'{o}ria no caso da luz. Se considerarmos a primeira vis\~{a}o, como o f\'{o}ton n\~{a}o tem massa, a rela\c{c}\~{a}o entre sua energia ($E$) e seu \textit{momentum} ($p$), dada pela f\'{o}rmula de Einstein, se escreve
$$ E = pc$$
Agora, considerando a luz como onda, seu comprimento de onda ($\lambda$) e sua frequ\^{e}ncia ($\nu$) satisfazem a rela\c{c}\~{a}o cl\'{a}ssica
$$ \lambda \nu = c$$
na qual $c$ \'{e} a velocidade de propaga\c{c}\~{a}o da luz. Por outro lado, a energia do f\'{o}ton, de acordo com a F\'{\i}sica Qu\^{a}ntica, \'{e} dada pela equa\c{c}\~{a}o de Planck
$$E = h \nu$$
enquanto seu \textit{momentum}, de acordo com de Broglie, vale
$$p = \frac{h}{\lambda}$$
As duas \'{u}ltimas equa\c{c}\~{o}es ``misturam'' as caracter\'{\i}sticas cl\'{a}ssicas de part\'{\i}cula (representadas pelas grandezas $E$ e $p$) com a ondulat\'{o}ria (caracterizada por $\nu$ e $\lambda$). Essa no\c{c}\~{a}o de dualidade onda-part\'{\i}cula s\'{o} se manifesta, portanto, em sistemas para os quais a a\c{c}\~{a}o envolvida \'{e} da ordem de $\hbar$. Note que, estritamente no caso de part\'{\i}culas de massa de repouso nula, como \'{e} o caso do f\'{o}ton, esse conjunto de quatro equa\c{c}\~{o}es \'{e} perfeitamente coerente. Ou seja, eliminando qualquer uma delas, a equa\c{c}\~{a}o eliminada \'{e} reobtida a partir das outras tr\^{e}s. No caso do el\'{e}tron, por exemplo, isso n\~{a}o \'{e} mais verdade. O mesmo ocorre no caso n\~{a}o relativ\'{\i}stico no qual a part\'{\i}cula livre tem uma energia $E= p^2/(2m)$, que n\~{a}o \'{e} mais linear em $p$.

\section{A coer\^{e}ncia de tudo: a Mec\^{a}nica Qu\^{a}ntica}
\paragraph*{}

Em \'{u}ltima an\'{a}lise, tanto Heisenberg quanto Schr\"{o}dinger buscaram inspira\c{c}\~{o}es no sucesso do modelo at\^{o}mico de Bohr. Como disse Einstein, deve haver algo fundamental por tr\'{a}s de suas ideias, j\'{a} que o modelo descreve muito bem a regularidade espectral dada pela f\'{o}rmula de Balmer. Entretanto, os pontos de partida foram t\~{a}o radicalmente distintos que determinaram dois formalismos inicialmente distintos e muito diferentes da Mec\^{a}nica Qu\^{a}ntica.

O fulcro da ideia de Heisenberg de 1925 \'{e} que a nova teoria deve se concentrar em quantidades observadas:
\begin{quotation}\baselineskip=10pt {\small
\noindent \textit{Agora, as coisas que voc\^{e} observa s\~{a}o apenas muito remotamente conectadas com as \'{o}rbitas de Bohr. Ent\~{a}o, Heisenberg disse que as \'{o}rbitas de Bohr n\~{a}o s\~{a}o muito im\-portantes. As coisas que s\~{a}o observadas, ou que est\~{a}o estreitamente conectadas com as quan\-tidades observadas, est\~{a}o todas associadas com duas \'{o}rbitas de Bohr e n\~{a}o apenas com uma \'{o}rbita de Bohr: \textit{duas} e n\~{a}o \textit{uma}.}}
\end{quotation}
\noindent o que, em \'{u}ltima an\'{a}lise, levou-o a descrever as grandezas f\'{\i}sicas como operadores matriciais, que n\~{a}o mais necessariamente comutam. Do ponto de vista epistemol\'{o}gico, a formula\c{c}\~{a}o de Heisenberg traz uma novidade inesperada: seu conhecido \textit{princ\'{\i}pio de incerteza}.

Este ponto de vista de Heisenberg n\~{a}o era t\~{a}o estranho aos pressupostos de Bohr, doze anos antes, quando ele baseou seu modelo at\^{o}mico em duas hip\'{o}teses: o \textit{postulado da frequ\^{e}ncia} e o \textit{princ\'{\i}pio de correspond\^{e}ncia da frequ\^{e}ncia}. Sua op\c{c}\~{a}o parece ser bem mais filos\'{o}fica que ditada por quest\~{o}es f\'{\i}sicas. O argumento apresentado a seguir \'{e} uma forma alternativa para estabelecer a quantiza\c{c}\~{a}o tanto da energia quanto do momento angular do el\'{e}tron em um \textit{estado estacion\'{a}rio} sem passar por outros. Talvez, uma explora\c{c}\~{a}o desse racioc\'{\i}nio, naquela circunst\^{a}ncia, poderia conduzir a uma vers\~{a}o da regra de quantiza\c{c}\~{a}o de Wilson-Sommerfeld, tal como sugerido na primeira palestra que Bohr deu em homenagem ao f\'{\i}sico dinamarqu\^{e}s Christian Christiansen, publicada em 1918. De fato, em poucas palavras, foi nesse artigo que Bohr mostrou o resultado agora bem conhecido de que a lei de quantiza\c{c}\~{a}o de Planck para a energia de um oscilador harm\^{o}nico unidimensional \'{e} equivalente \`{a} condi\c{c}\~{a}o
$$
\oint p \mbox{d} q = n h
$$
em que a integral \'{e} feita sobre uma oscila\c{c}\~{a}o completa da vari\'{a}vel $q$ entre seus limites e $p$ \'{e} o \textit{momentum} canonicamente conjugado.

O f\'{\i}sico dinamarqu\^{e}s chegou a essa conclus\~{a}o ap\'{o}s ter percebido o quanto a \textit{hip\'{o}tese adiab\'{a}tica}, introduzida por Ehrenfest, que a chamou de princ\'{\i}pio de transformabilidade mec\^{a}nica, poderia dar suporte \`{a} sua de\-fi\-ni\c{c}\~{a}o de \textit{\'{o}rbitas estacion\'{a}rias}, ou, em outras palavras, como ele justificaria a fixa\c{c}\~{a}o de uma s\'{e}rie de estados at\^{o}micos \textit{entre a cont\'{\i}nua multitude de poss\'{\i}veis movimentos mec\^{a}nicos}.

E, como visto na Se\c{c}\~{a}o~\ref{de_Broglie}, de Broglie mostra que a coer\^{e}ncia com essa \'{o}rbita estacion\'{a}ria vai requerer que $p = h/\lambda$. Nesse ponto cabe o seguinte coment\'{a}rio. Como ressalva o f\'{\i}sico russo Dmitri Ivanovich Blokhintsev, se, nesta rela\c{c}\~{a}o de Louis de Broglie, entende-se por $\lambda$ um comprimento de onda, n\~{a}o faz sentido sustentar que uma part\'{\i}cula se encontra em uma posi\c{c}\~{a}o definida, uma vez que o comprimento de onda \'{e}, por defini\c{c}\~{a}o, a caracter\'{\i}stica de uma onda plana monocrom\'{a}tica que pressup\~{o}e uma extens\~{a}o que se repete periodicamente no infinito espacial $(-\infty \leq x \leq \infty)$, algo, portanto, espacialmente extenso. Portanto, seria a rela\c{c}\~{a}o de L.~de Broglie uma pista para o princ\'{\i}pio de incerteza de Heisenberg?

Por outro lado, como \'{e} bem sabido, o caminho de Schr\"{o}dinger na constru\c{c}\~{a}o da Mec\^{a}nica Qu\^{a}ntica originou-se com as ideias de Louis de Broglie, que o levaram \`{a} sua famosa equa\c{c}\~{a}o. Schr\"{o}dinger recebeu um convite para trabalhar na Universidade de Z\"{u}rich, em um grupo de pesquisas liderado por Debye. Ao ler os trabalhos de L. de Broglie sobre a \textit{onda de mat\'{e}ria}, Debye sugeriu a Schr\"{o}dinger que fizesse um semin\'{a}rio sobre as ideias do pr\'{\i}ncipe franc\^{e}s. O primeiro impulso de Schr\"{o}dinger foi recusar, dizendo: -- \textit{Eu n\~{a}o quero falar sobre tal ``nonsense''}. Por\'{e}m, como Debye era o chefe do grupo de pesquisa, o f\'{\i}sico austr\'{\i}aco teve consci\^{e}ncia de que esse semin\'{a}rio era relevante para a forma\c{c}\~{a}o do referido grupo. Aceitou, portanto, a sugest\~{a}o e prometeu apresentar a contribui\c{c}\~{a}o de Louis de Broglie em uma forma matem\'{a}tica mais compreens\'{\i}vel. Foi assim que, em um semin\'{a}rio para o grupo, prop\^{o}s sua c\'{e}lebre equa\c{c}\~{a}o. Por ocasi\~{a}o da apresenta\c{c}\~{a}o desse semin\'{a}rio,\footnote{\, Conforme Debye contou ao f\'{\i}sico russo Pyotr Leonidovich Kapitza.} Schr\"{o}dinger n\~{a}o estava demasiadamente convicto da equa\c{c}\~{a}o que estava propondo. Entretanto, Debye, presente a esse semin\'{a}rio, disse a Schr\"{o}dinger, ao t\'{e}rmino de sua palestra: -- \textit{Voc\^{e} fez um trabalho extraordin\'{a}rio}~\cite{Kapitza}.

Explorando uma equival\^{e}ncia formal entre o princ\'{\i}pio do tempo m\'{\i}nimo de Fermat, na \'{O}ptica, e o de m\'{\i}nima a\c{c}\~{a}o de Hamilton para a Mec\^{a}nica Cl\'{a}ssica, Schr\"{o}dinger, que era um estudioso de fen\^{o}menos \'{o}pticos, vai explorar essa semelhan\c{c}a, conhecida como \textit{Analogia \'{O}ptico-Mec\^{a}nica}, para encontrar uma equa\c{c}\~{a}o diferencial que descrevesse a onda de mat\'{e}ria de Louis de Broglie. Para surpresa de todos, a solu\c{c}\~{a}o geral da dita equa\c{c}\~{a}o de Scrh\"{o}dinger dependente do tempo, n\~{a}o pode ser uma fun\c{c}\~{a}o nem real, nem imagin\'{a}ria pura: \'{e} necessariamente complexa. Este fato levou, como todos sabem, Max Born a propor uma interpreta\c{c}\~{a}o probabil\'{\i}stica para a nova Mec\^{a}nica Qu\^{a}ntica. Mas isso \'{e} outra hist\'{o}ria...

\section{Coment\'{a}rios finais}
\paragraph*{}

A hist\'{o}ria de uma crise sem precedentes, aqui resumida, resultou em uma nova forma de se ver e descrever o microcosmo. Ela talvez n\~{a}o pudesse ter sido escrita, ou seria apenas parcialmente concebida, se f\'{\i}sicos do porte de Langevin, Einstein e Born n\~{a}o tivessem tido o discernimento para perceber o valor das ``ideias estranhas'' de Louis de Broglie. AO leitor interessado em mais detalhes sobre sua contribui\c{c}\~{a}o \`{a} F\'{\i}sica sugere-se consultar, por exemplo, o Cap\'{\i}tulo 5 da Ref.~\cite{Golub}.

\section*{Agradecimentos}
\paragraph*{}

Os autores agradecem a Eduardo Sim\~{o}es pelo convite para participar do Ciclo de Encontros sobre Louis de Broglie, assim como aos colegas pelo ambiente acolhedor e prop\'{\i}cio ao debate de ideias e, por fim, pela oportunidade de contribuir com o presente texto para o volume 2 do livro \textit{Filosofia dos F\'{\i}sicos}.

\vspace{0.5cm}

\centerline{------------------------------------------------------------------------------------------------------------------}

\end{document}